\documentclass{aa}  
\usepackage[dvips]{graphicx}
\usepackage{txfonts}
\usepackage{natbib}
\bibpunct{(}{)}{;}{a}{}{,} % to follow the A&A style
\begin{document}
   \title{Mid-infrared VISIR and \textit{Spitzer} observations of the
     surroundings of the magnetar SGR~1806-20\thanks{Based on
  observations carried out at the European Southern Observatory
  under programmes ID 075.D-0773 and 077.D-0721 (P.I. S. Chaty).}}

  \author{F. Rahoui
    \inst{1,2}
    \and
    S. Chaty\inst{1}
    \and
    P-O. Lagage\inst{1}	
  }

  \offprints{F. Rahoui}

  \institute{Laboratoire AIM, CEA/DSM - CNRS - Universit\'e Paris Diderot,
    IRFU/Service d'Astrophysique, B\^at. 709, CEA-Saclay, F-91191
    Gif-sur- Yvette C\'edex, France, \email{frahoui, chaty, pierrre-olivier.lagage@cea.fr}
      \and
      European Southern Observatory, Alonso de C\'ordova 3107,
      Vitacura, Santiago de Chile\\}
   \date{Received; accepted}

% \abstract{}{}{}{}{} 
% 5 {} token are mandatory
 
   \abstract
       {SGR~1806-20 is the soft
         gamma-ray repeater that 
         exhibited the highest activity since its
         discovery, including a giant flare in 2004, December
         27. Previous studies of this source - which is likely a
         magnetar - showed that it was associated with a
         massive star cluster embedded into a gas and dust cloud.
         Moreover, several stars in the cluster are
         peculiar hypergiants - luminous blue variable (LBV) and Wolf-Rayet
         stars - exhibiting strong and likely dusty
         stellar winds.}	 
       {We aimed at studying the mid-infrared emission of the stars
         associated with the same cluster as SGR~1806-20, to detect
         variations that could be due to the high-energy
         activity of the magnetar through interaction with the dust. 
         We also aimed at studying the morphology of the cloud
         close to the centre of the cluster.}
       {We carried out mid-infrared observations of SGR~1806-20 and
         its environment - with the highest spatial resolution in this domain 
         to date - using ESO/VISIR in 2005 and 2006, and we 
         retrieved \textit{Spitzer}/IRAC-MIPS archival data of the same
         field. We performed broadband photometry of three stars
         - LBV~1806-20, a WC9 and
	 an O/B supergiant - on our VISIR images, as well as on the IRAC
         data. We then built and fitted their broadband
         spectral energy distributions with
	 a combination of two absorbed black bodies, representing their
	 stellar components, as well as a possible mid-infrared excess,
         in order to derive their physical parameters.}
       {We show that LBV~1806-20 and the WC9 star exhibit a
         mid-infrared excess, likely because of the presence of
         circumstellar dust
         related to their winds. We also show that only LBV~1806-20 had a
         variable flux over a period of two
	 years, variability which is due to its LBV nature
         rather than to a heating of the gas and dust cloud by the
         high-energy emission of SGR~1806-20. 
         Finally, differences in the intrinsic absorptions of the
         three stars show an inhomogeneous structure of the density of
         the gas and dust cloud in the massive star cluster.}
       {}
       
   
   \maketitle
%
%________________________________________________________________

\section{Introduction}

Soft gamma-ray repeaters (SGRs) represent a small group - four known
objects, three in the Galactic centre and one in the Large Magellanic Cloud - of
highly-magnetized exotic pulsars presenting no evidence of
binarity. They are characterised by a spin period clustered in the range
6-12~s, and a persistent soft X-ray emission around
$10^{35}-10^{36}$~erg~$\textrm{s}^{-{1}}$. Their spectra are
well-fitted by a combination of a black body (kT$\sim$0.5-0.6~keV) and
a power-law tail ($\Gamma\sim2.5$), and they also exhibit short and intense bursts
(duration of a few hundred milliseconds) of soft
$\gamma$-rays and hard X-rays. Giant flares 
have also been observed for three of them: SGR~0526-66, SGR~1806-20 and
SGR~1900+14. Recently, persistent hard X-ray emission has been detected on 
SGR~1806-20 \citep{2005Mereghetti}.
 
As for anomalous X-ray pulsars (AXPs) - proposed to be strongly
related to SGRs due to their similar characteristics \citep{2002Gavriil} - the likely
source of their radiative emission is not accretion like
for other neutron stars but their very strong magnetic field
($10^{14}\,-\,10^{15}$~G) \citep[see e.g.][for recent reviews]{2002Mereghetti, 2006Woods}.
It is now commonly accepted that SGRs and AXPs are 
magnetars, i.e. neutron stars whose high-energy emission (persistent and
transient) is powered by the decay
of a very strong dipole-like magnetic
field \citep{1992Duncan, 1995Thompson, 1996Thompson}.

SGR~1806-20 is the SGR that exhibited the highest activity since its
discovery \citep{1987Atteia, 1987Laros}. It is located towards the
Galactic centre, exhibits 7.47~s pulsations
\citep{1998Kouveliotou} and an 8.3$\,10^{-{11}}$~s~s$^{-{1}}$
spin-down rate. It entered in an active phase in 2003 and exhibited bursts up to
late 2004. Finally, a giant flare (corresponding to $10^{47}-10^{48}$~erg~s$^{-1}$ for a distance between 8 and 15~kpc) occurred in 2004 December 27 \citep{2005Hurley}, and a remnant radio afterglow allowed an accurate
localisation of the source \citep{2005Gaensler}. 

LBV~1806-20, a luminous blue variable (LBV), was long
believed to be its near-infrared (NIR) counterpart
\citep{1995Kulkarni}, but \citet{2002Kaplan} showed that the SGR was
12$\arcsec$ away from the LBV using \textit{Chandra} observations. The accurate localisation of the
source, as well as an exhaustive
monitoring finally led to the detection of the
NIR counterpart of SGR~1806-20 \citep{2005Israel, 2005Kosugi} by
observing a variability correlated to
the high-energy emission, as for AXPs, which strengthens the connection
between both objects.

The environment of SGR~1806-20 is very young and dusty, and it is then
relevant to observe it at mid-infrared (MIR) wavelengths and search for variabilities correlated to the
high-energy emission. Indeed, \citet{1999Fuchs} and
\citet{2001Eikenberry} showed that it was associated with a massive star
cluster, along with LBV~1806-20. Moreover, this cluster is located
inside the radio nebula G10.0-0.3, powered by the very strong winds of
the LBV \citep{2001Gaensler}. Finally, this radio nebula itself is
associated with the giant molecular complex W31
\citep{1997Corbel}. \citet{2004Corbel} showed that W31 was resolved into two components, one
at $d\sim4$~kpc with $A_{\rm v}\sim15$, and another one at
$d\sim15.1$~kpc with $A_{\rm v}=37\pm3$,
and that LBV~1806-20 (and
consequently the massive star cluster, and SGR~1806-20) belonged to the
further component of W3, suggesting a distance of $15.1_{-1.3}^{+1.8}$~kpc. 
Nevertheless, \citet{2008Bibby} recently performed high-resolution near-infrared spectroscopy of O/B and 
Wolf-Rayet stars in the cluster to derive their accurate spectral classification. This allowed them - 
using both synthetic photometry and isochrone fitting - to derive a distance modulus of $14.7\pm0.35$~mag 
for the cluster, which corresponds to a distance of $8.7_{-1.5}^{+1.8}$~kpc.
\newline

In this paper, we report MIR observations - with the highest spatial resolution in this domain to date - 
of SGR~1806-20 and its environment, carried out at ESO/VLT with VISIR in June 2005 and 2006. We also report archival data obtained at three
different epochs between 2004 and 2006 with IRAC and MIPS mounted on
\textit{Spitzer}. Our goal was to detect any MIR
variability which could be due to the high-energy activity of
the SGR, as well as to study the morphology of the gas and dust cloud in
which all sources are embedded.

In section 2, we describe these observations and their analysis. In
section 3, we present the broadband spectral energy distributions
(SEDs) we built for three sources 
using the VISIR and IRAC fluxes obtained in 2005 and 2006. For each
year, there were no more than three months between the VISIR and the
IRAC observations, which allowed us to fit these SEDs to
derive the physical parameters of the stars with contemporaneous
data from 3.6~$\mu$m to 11.25~$\mu$m. In section 4, we analyse the
results and discuss the MIR variability, the intrinsic
absorption, and the distance of these stars, as well as the MIR emission of
SGR~1806-20.

\begin{figure*}
  \centering
  \includegraphics[height=5cm,width=5cm]{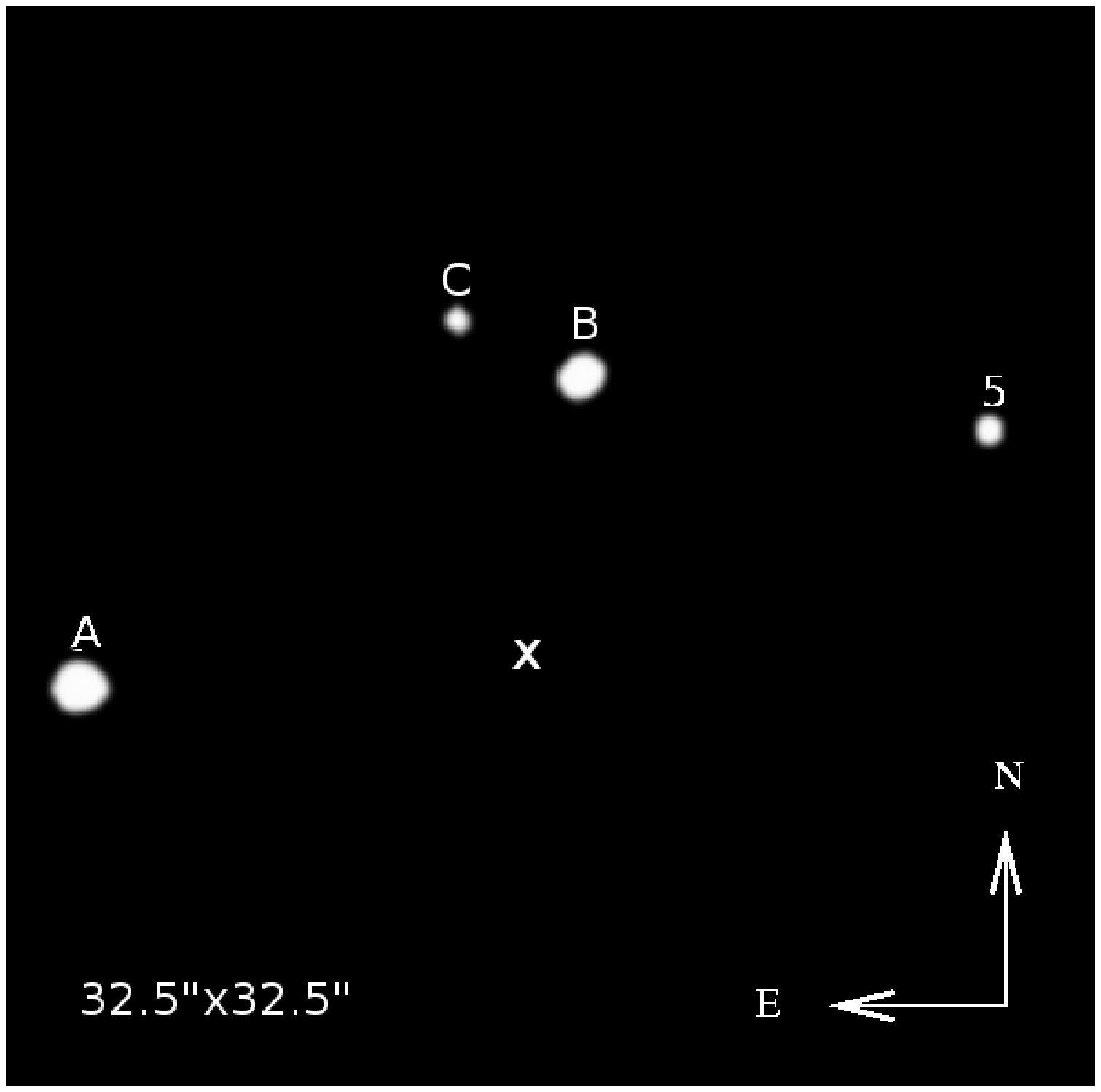}
  \includegraphics[height=5cm,width=5cm]{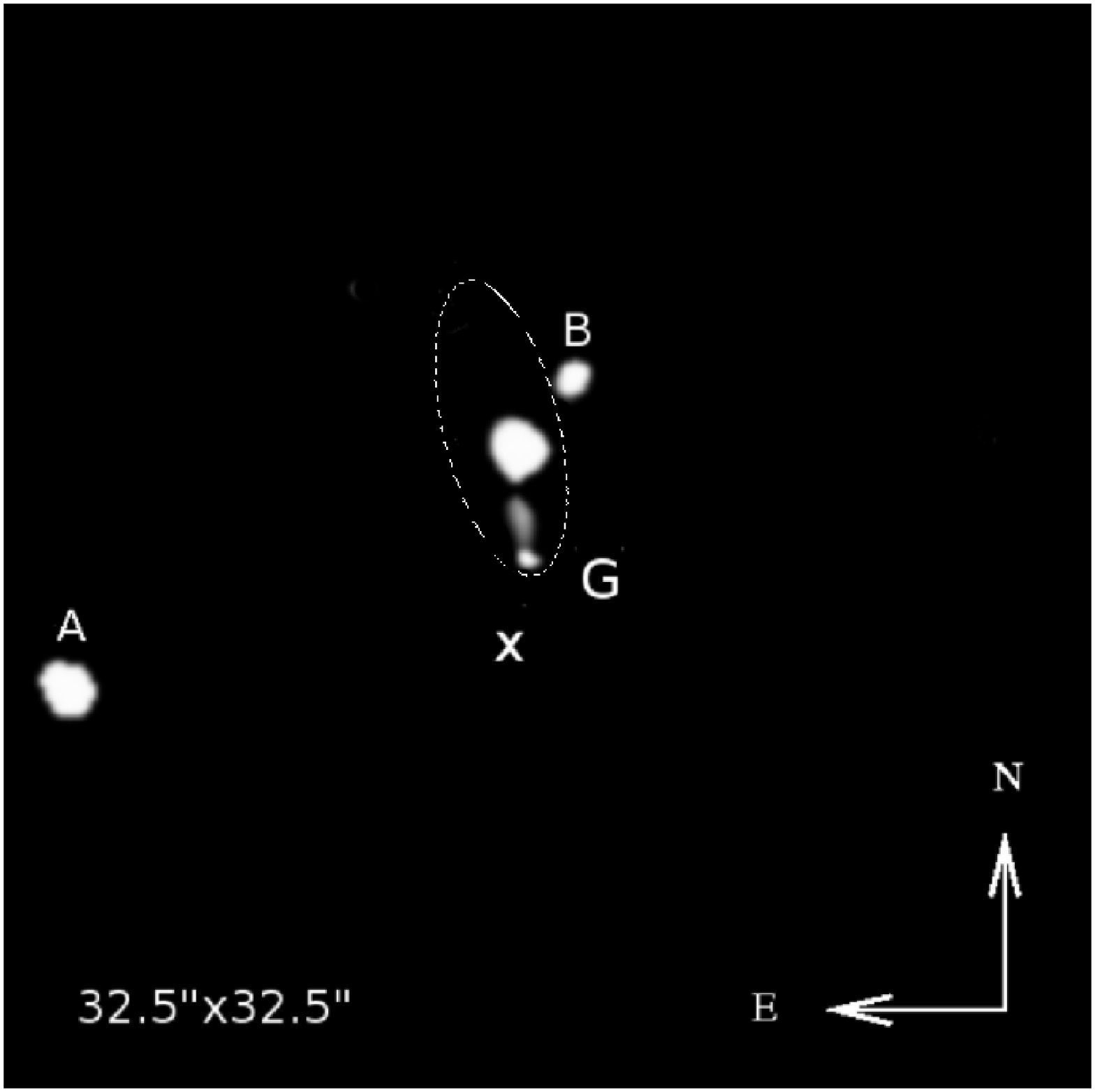}
  \includegraphics[height=5cm,width=5cm]{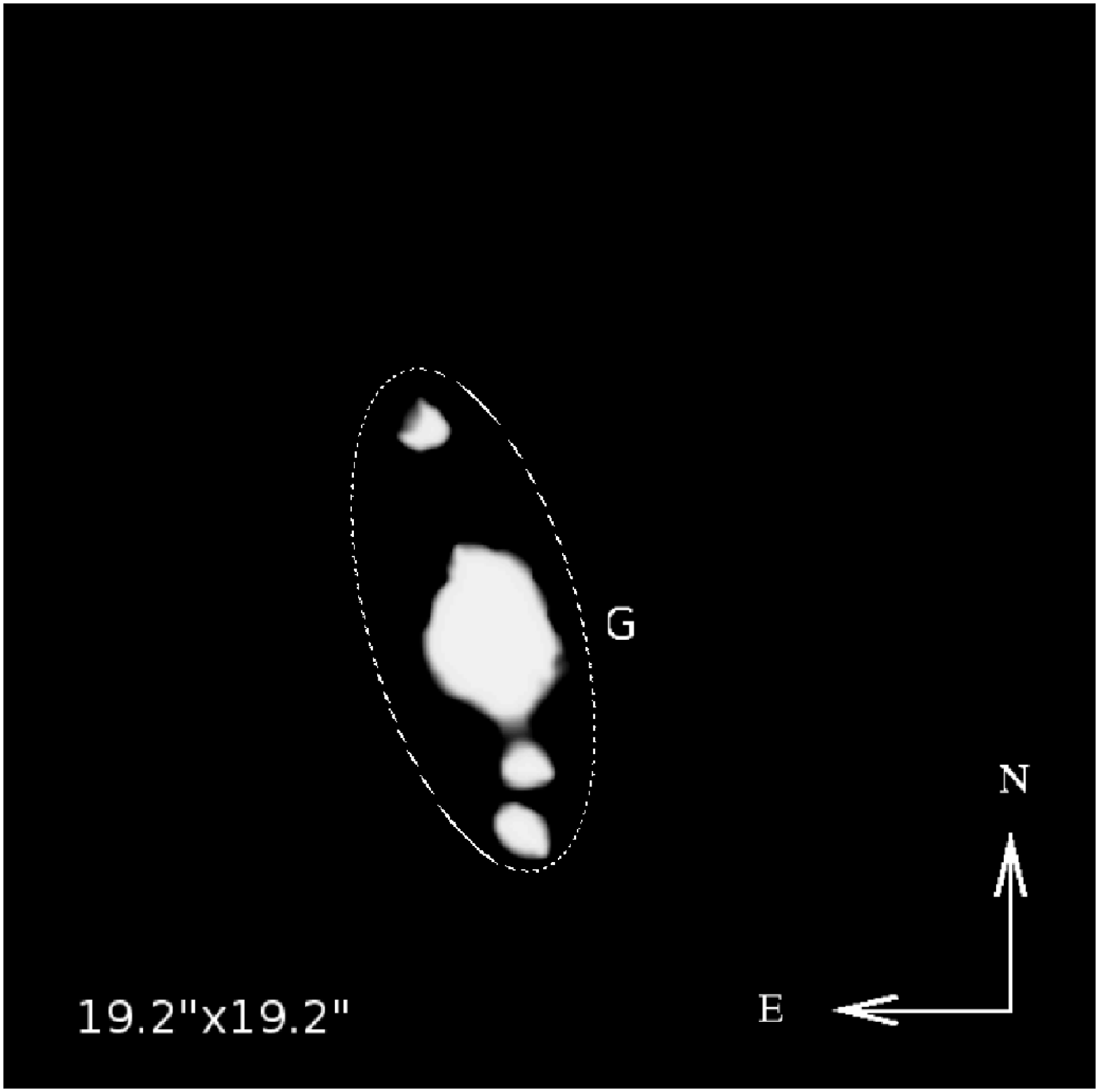}
  \caption{PAH1 (left, 0\farcs127 plate scale per pixel), PAH2
  (centre, 0\farcs127 plate scale per pixel), and Q2 (right,
  0\farcs075 plate scale per pixel) VISIR images of the environment of SGR~1806-20
  (white X point) obtained in June 2006. A is LBV~1806-20, B is a WC9 star, C is an O/B
  supergiant, and 5 is a Red Giant \citep{2004Eikenberry,2005Figer}. 
  An ellipse pattern surrounds G, the hottest part of the gas and 
  dust cloud in which all stars are embedded. We detected it in PAH2 and in Q2.}
\end{figure*}

\begin{figure*}
  \centering
  \includegraphics[height=4cm,width=7cm]{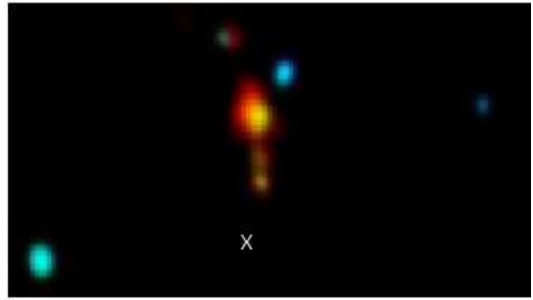}
  \caption{Three-color image of the environment of SGR~1806-20 (white X
  point) done with images of June 2006; PAH1 is in blue, PAH2 in green,
  and Q2 in red. Q2 image was rescaled to a 0\farcs127 plate scale. The cloud
  detected with VISIR appears to be splitted into several
  components, which shows that its temperature is not homogeneous, the
  cloud being hotter in the centre of the massive star cluster.}
\end{figure*}

\section{Observations and data reduction}

\subsection{ESO/VISIR data}

The MIR observations were carried out on 2005 June 20-22 and 2006 June
29-30 using VISIR \citep{2004Lagage}, 
the ESO/VLT mid-infrared imager and spectrograph, composed of an
imager and a long-slit spectrometer covering several filters in
\textit{N} and \textit{Q} bands and
mounted on Unit 3 of the VLT (Melipal). The standard ``chopping
and nodding'' MIR observational technique was used to suppress the
background dominating at these wavelengths. Secondary mirror-chopping was
performed in the north-south direction with an amplitude of
16$\arcsec$ at a frequency of 0.25~Hz. Nodding technique,
needed to compensate for chopping residuals, was chosen as
parallel to the chopping and applied
using telescope offsets of 16$\arcsec$. Because of the high
thermal MIR background for ground-based
observations, the detector integration
time was set to 16~ms.

We performed broadband photometry in three filters, PAH1
($\lambda$=8.59$\pm$0.42~$\mu$m), PAH2
($\lambda$=11.25$\pm$0.59~$\mu$m), and Q2
($\lambda$=18.72$\pm$0.88~$\mu$m), using the large field in all bands 
(32\farcs5$\times$32\farcs5$ $ and 0\farcs127 plate scale per pixel) in June
2005. In June 2006, we used the large field in PAH1 and PAH2 and the
small one (19\farcs2$\times$19\farcs2 and 0\farcs075 plate scale per pixel)
in Q2. All the observations were bracketed with standard star
observations for flux calibration and
PSF determination. The weather conditions
were good and stable during the observations.

Raw data were reduced using the IDL reduction package written by E.
Pantin. The elementary images were co-added in real-time to obtain
chopping-corrected data, then the different nodding positions were
combined to form the final image. The VISIR detector is affected by
stripes randomly triggered by some abnormal high-gain pixels.
A dedicated destriping method was developed (Pantin 2008,
in prep.) to suppress them.

The filtered reduced images of 2006 are displayed in Fig~1. These images
were cleaned using the multiresolution software package written by
J.L. Starck\footnote{http://thames.cs.rhul.ac.uk/$\sim$multires/mr4-software/index.html} and presented in \citet{1998Starck}, which performs
background and noise modelling, as well as
noise subtraction using multiresolution tools. Nevertheless, we
performed photometry on the unfiltered images using aperture
photometry, and the fluxes in all
bands are listed in Table~1. We kept the same labels as \citet[][EML04
  hereafter]{2004Eikenberry} and \citet{2005Figer} for all the detected stars. A is
therefore LBV~1806-20, B is a WC9 star, C
is an O/B supergiant, and 5 is a Red Giant. Finally, G is a hot and
dense part of the gas and
dust cloud which is the likely birth site of the massive star cluster,
and is only visible in the MIR domain. On June 2005, the exposure time
was 600~s in all bands while it was 1200~s in PAH1 and PAH2 and 2400~s
in Q2 in June 2006, which explains why we did not detect neither C in
PAH1 nor G in PAH2 in June
 2005, their fluxes being too low compared to the sensitivity
 we were able to reach (see Table~1). A three-color image (made with PAH1, PAH2, and Q2 images) is also displayed in Fig~2 and we clearly see that
the gas and dust cloud emits mostly beyond 11~$\mu$m.

\begin{figure*}
\centering
\begin{tabular}{cc}
  \includegraphics[width=6cm,height=4.5cm]{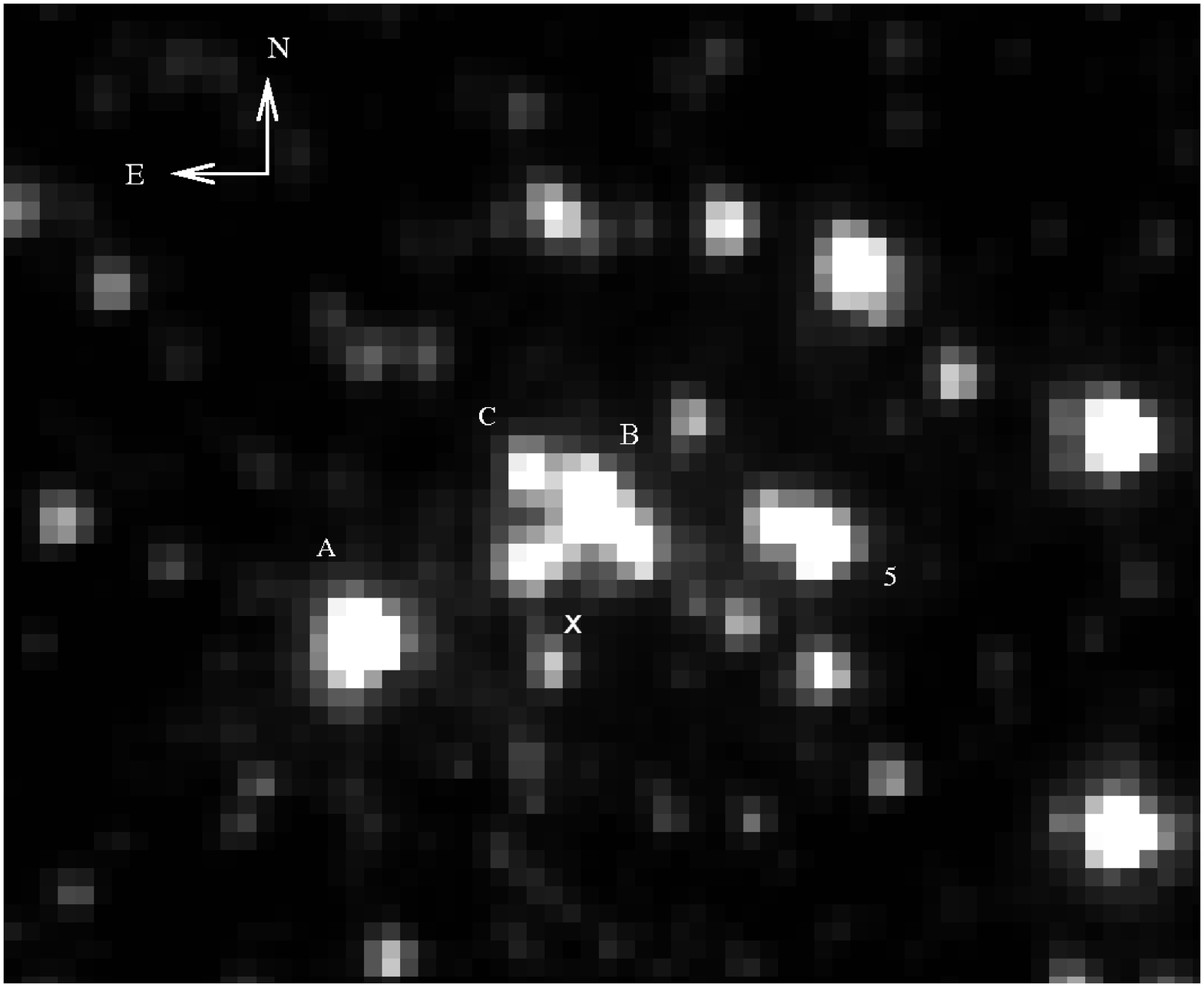}&\includegraphics[width=6cm,height=4.5cm]{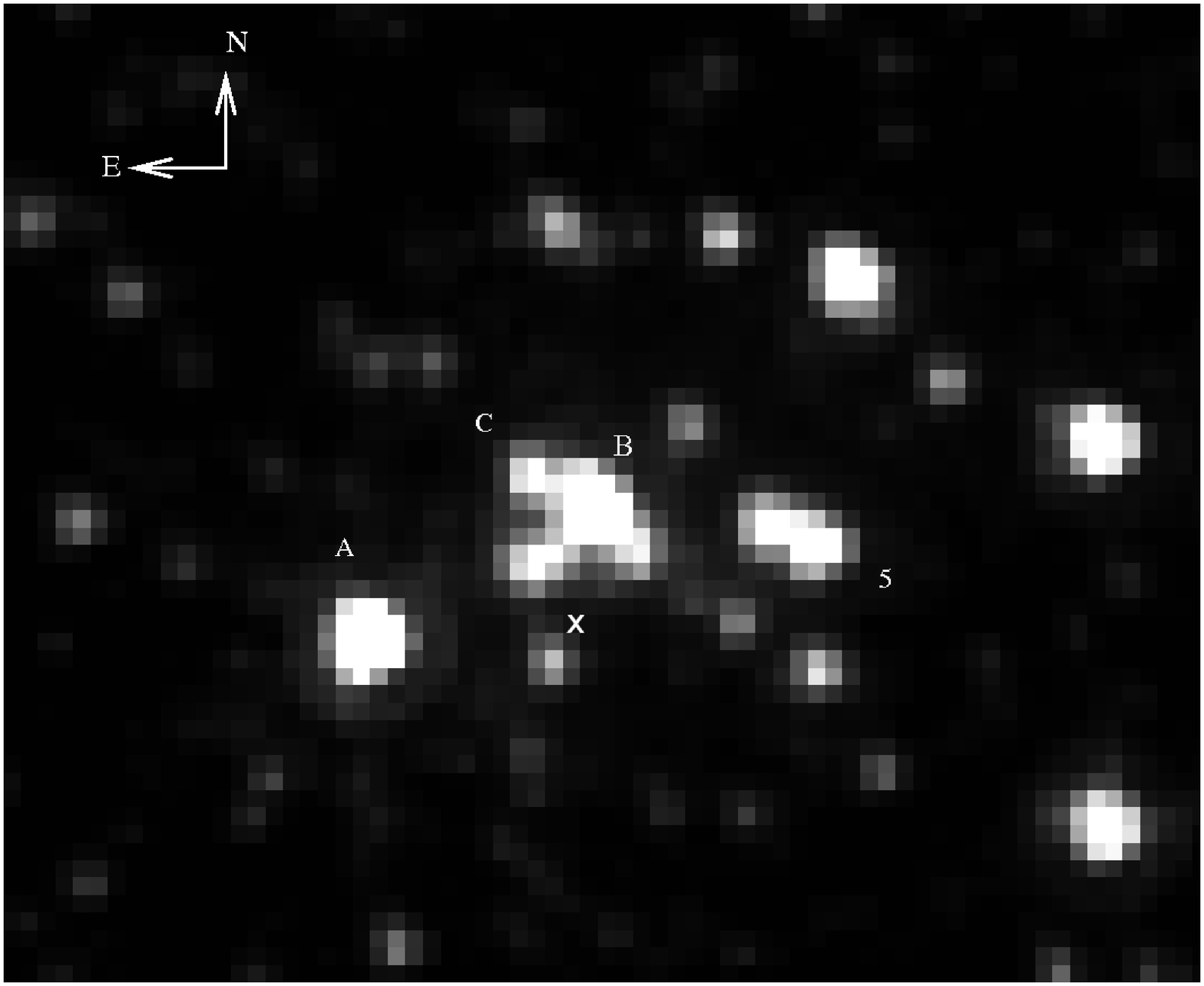}\\
  \includegraphics[width=6cm,height=4.5cm]{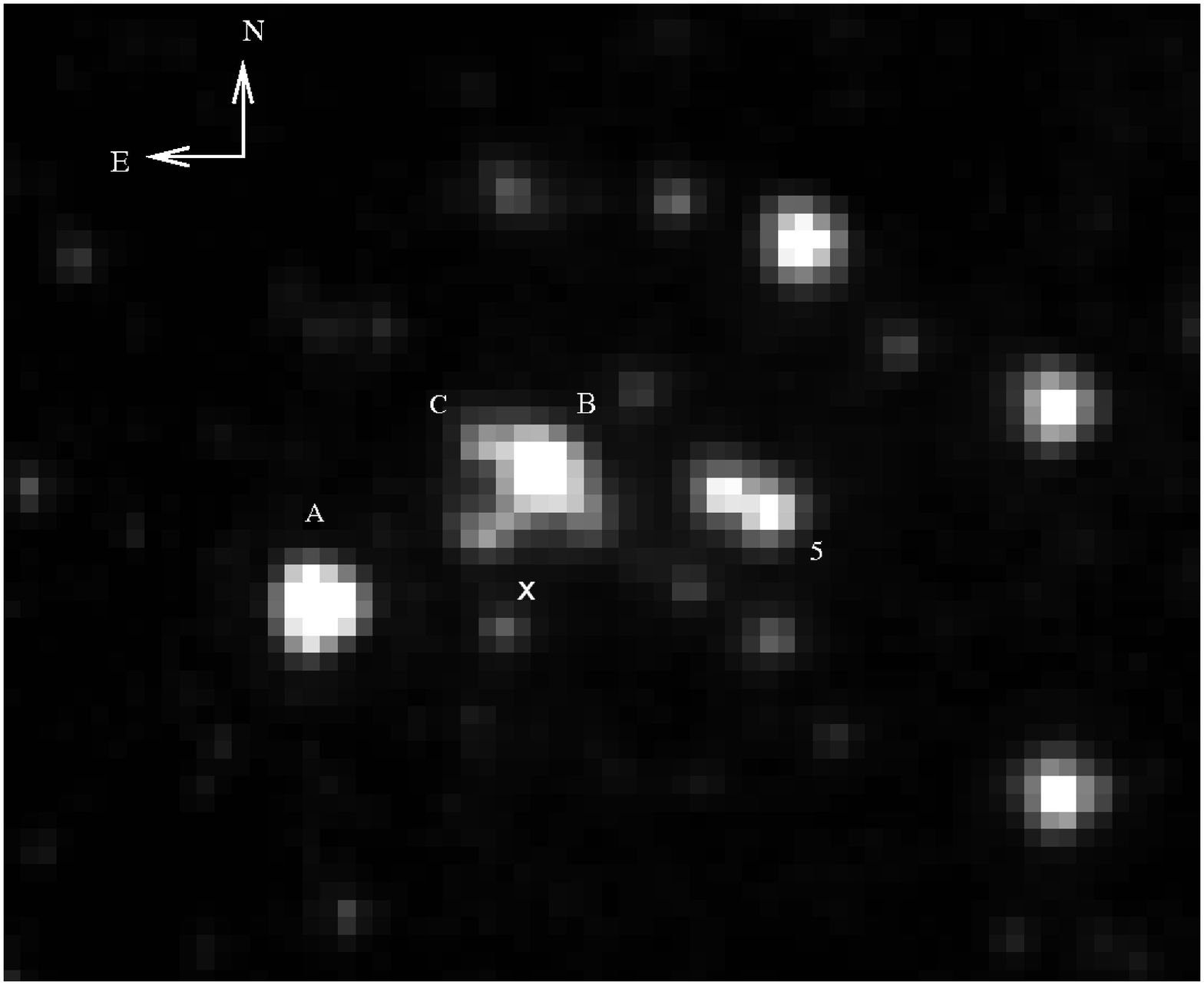}&\includegraphics[width=6cm,height=4.5cm]{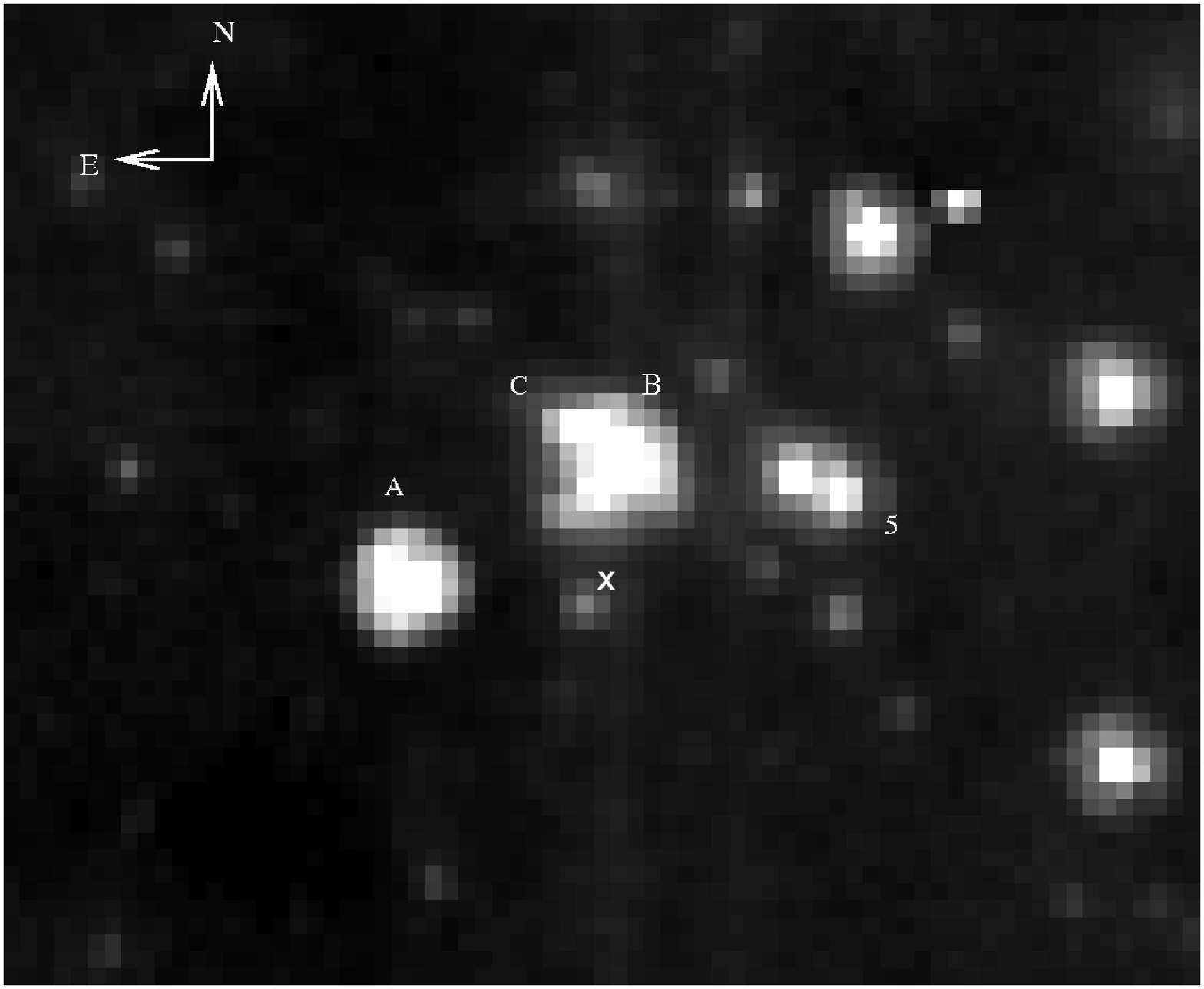}\\
\end{tabular}
  \caption{IRAC 3.6~$\mu$m (top left), 4.5~$\mu$m (top right), 5.8
   ~$\mu$m (bottom left) and 8.0~$\mu$m (bottom right) images of
  the environment of SGR~1806-20 (white X point). The field of view is about
  79$\arcsec$$\times$71$\arcsec$ and the plate scale is 1\farcs2 per pixel.
  We  point out that the position of SGR~1806-20 is 2\farcs95 away from the point source located i south-east of the white X point, which - considering the 0\farcs5 astrometric accuracy of the \textit{Spitzer} telescope - excludes an association with 
  SGR~1806-20.}
\end{figure*}

\begin{figure*}
  \centering
  \includegraphics[height=6cm]{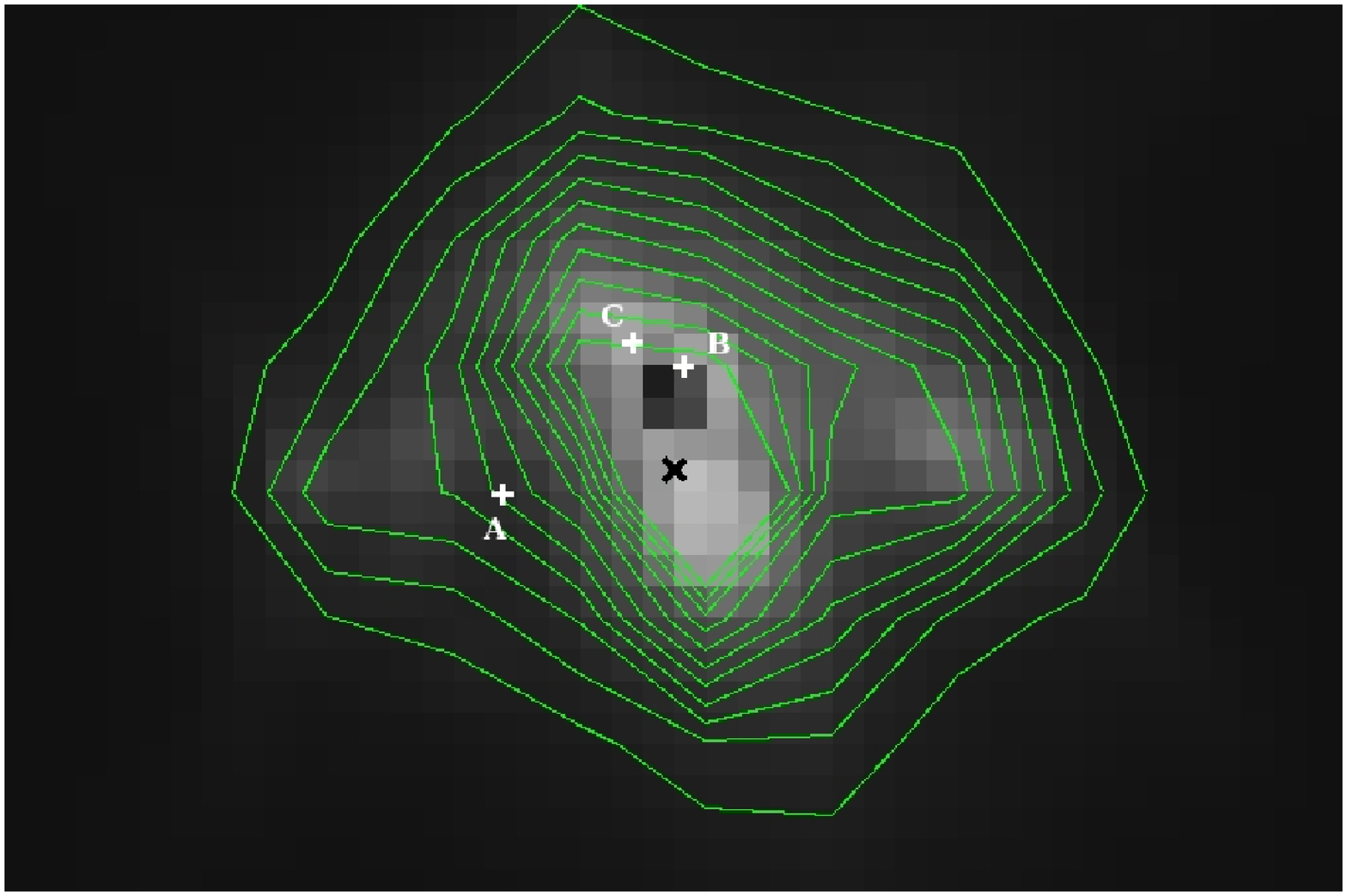}
  \caption{MIPS 24~$\mu$m image of the environment of SGR~1806-20
    (black X point). We also superimposed the MIPS contour plot
  to stress the  differences in luminosity. We clearly see the gas and
  dust cloud in which all the
  stars are embedded, and which is likely the birth site of the massive
  star cluster. It appears that LBV~1806-20 (A) is located in a
  less dense and colder zone of the cloud than the WC9 star (B) and
  the O/B supergiant star (C), which
  explains that its intrinsic absorption is lower. Moreover, the central
  part of the cloud - which appears saturated in this image as it is too hot
  with respect to the MIPS level of saturation - corresponds to the
  region detected with VISIR. The field of view is
  94$\arcsec$$\times$67$\arcsec$ and the plate scale is 2\farcs5 per pixel.}
\end{figure*}

\subsection{\textit{Spitzer}/IRAC-MIPS archival data}

We searched for MIR public data of the star cluster in the archives of
the \textit{Spitzer space telescope}. We found
photometric data taken at three different epochs, 2004 October 29,
2005 September 27, and 2006 April 10, in the \textit{Spitzer}'s Galactic Legacy Infrared Mid-Plane Survey
Extraordinaire \citep[GLIMPSE,][]{2003Benjamin}, survey of the Galactic plane
($|b|\,\leq\,1^\circ$ and between \textit{l}=10$^\circ$ and
\textit{l}=65$^\circ$ on both sides of the Galactic centre), using the IRAC
camera in four bands,  3.6$\pm$0.745~$\mu$m, 4.5$\pm$1.023~$\mu$m,
5.8$\pm$1.450~$\mu$m and 8.0$\pm$2.857~$\mu$m. IRAC plate scale is
1\farcs2 per pixel and the field of view is 5\farcm2$\times$5\farcm2. We also found archival
MIPS data of the same field at 24~$\mu$m.

We performed photometry of the stars A, B and C we had detected with VISIR on
the post-Basic Calibration Data (post-BCD) using the software
\textit{StarFinder}, part of the \textit{Scisoft} package from ESO,
well-suited for point-source photometry
in crowded fields. Post-BCD data are raw data on which the
\textit{Spitzer} pipeline performs dark subtraction, multiplexer bleed
correction, detector linearization, flat-fielding, cosmic ray
detection, flux calibration, pointing refinement, mosaicking, and coaddition.
The images taken in September 2005 are displayed in Fig~3, and fluxes for
each epoch are listed in Table~2. Moreover, Fig~4 displays the MIPS
image of the same field at 24~$\mu$m.

\section{Absorption and SEDs}
\begin{table*}
  \caption{Summary of VISIR observations of the sources in the environment
	  of SGR~1806-20. We give their name and
	  their MIR fluxes (in Jy) in the PAH1 (8.59~$\mu$m), PAH2 (11.25~$\mu$m), and
	  Q2 (18.72~$\mu$m) filters. When we did not detect the
          source, we give its upper limit. For the dust cloud, 
         we give the upper limit for an extended source covering 450 pixels in 
         PAH1 and PAH2, as this corresponds to the area covered by the dust 
         cloud when we detected it in PAH2 in June 2006. Uncertainties are given at 1$\sigma$.}
  $$ 
  \begin{array}{cccc}
    \hline
    \textrm{Sources}&\textrm{PAH1}&\textrm{PAH2}&\textrm{Q2}\\
    \hline
    \hline
     &\multicolumn{3}{c}{\textrm{June 2005}}\\
    \hline
    \textrm{LBV~1806-20 (A)}&0.133\pm0.004&0.089\pm0.005&<0.058\\
    \hline
    \textrm{WC9 (B)}&0.077\pm0.003&0.057\pm0.004&< 0.058\\
    \hline
    \textrm{Supergiant (C)}&<0.0126&<0.0149&<0.058\\
    \hline
    \textrm{Red Giant (5)}&0.020\pm0.002&<0.0149&<0.058\\
    \hline
    \textrm{Dust (G)}&< 0.091&< 0.111&2.60\pm0.20\\
    \hline
    \hline
    &\multicolumn{3}{c}{\textrm{June 2006}}\\
    \hline
    \textrm{LBV~1806-20 (A)}&0.103\pm0.004&0.067\pm0.004&<0.051\\
    \hline
    \textrm{WC9 (B)}&0.076\pm0.003&0.053\pm0.003&<0.051\\
    \hline
    \textrm{Supergiant (C)}&0.012\pm0.002&< 0.0124 &<0.051\\
    \hline
    \textrm{Red Giant (5)}&0.019\pm0.002&< 0.0124&<0.051\\
    \hline
    \textrm{Dust (G)}&< 0.112&0.078\pm0.007&2.82\pm0.14\\
    \hline
  \end{array}
  $$ 
\end{table*}

\begin{table*}
  \caption{Summary of IRAC observations of the sources in
    the environment of SGR~1806-20. We give their name and
    their MIR fluxes (in Jy). Uncertainties are given at 1$\sigma$.}
  $$ 
  \begin{array}{ccccc}
    \hline
    \textrm{Sources}&3.6\,{\mu}\textrm{m}&4.5\,{\mu}\textrm{m}&5.8\,{\mu}\textrm{m}&8.0\,{\mu}\textrm{m}\\
    \hline
    \hline
    &\multicolumn{4}{c}{\textrm{October 2004}}\\
    \hline
    \textrm{LBV~1806-20 (A)}&0.306\pm0.010&0.316\pm0.008&0.283\pm0.008&0.175\pm0.005\\
    \hline
    \textrm{WC9 (B)}&0.165\pm0.005&0.208\pm0.006&0.218\pm0.007&0.141\pm0.003\\
    \hline
    \textrm{Supergiant (C)}&0.041\pm0.002&0.036\pm0.002&0.028\pm0.002&0.027\pm0.002\\
    \hline
    \hline
    &\multicolumn{4}{c}{\textrm{September 2005}}\\
    \hline
    \textrm{LBV~1806-20 (A)}&0.337\pm0.015&0.332\pm0.016&0.299\pm0.014&0.199\pm0.007\\
    \hline
    \textrm{WC9 (B)}&0.172\pm0.009&0.218\pm0.011&0.224\pm0.011&0.140\pm0.005\\
    \hline
    \textrm{Supergiant (C)}&0.037\pm0.003&0.037\pm0.003&0.031\pm0.003&0.027\pm0.001\\
    \hline
    \hline
    &\multicolumn{4}{c}{\textrm{April 2006}}\\
    \hline
    \textrm{LBV~1806-20 (A)}&0.297\pm0.010&0.304\pm0.001&0.272\pm0.011&0.174\pm0.005\\
    \hline
    \textrm{WC9 (B)}&0.170\pm0.006&0.207\pm0.007&0.225\pm0.009&0.140\pm0.006\\
    \hline
    \textrm{Supergiant (C)}&0.041\pm0.003&0.038\pm0.002&0.030\pm0.003&0.025\pm0.002\\
    \hline
  \end{array}
  $$ 
\end{table*}
\subsection{Absorption}
      
The absorption at wavelength $\lambda$, $A_{\rm \lambda}$, is a
crucial parameter to fit the SEDs, especially in the
MIR. Indeed, non accurate values can lead to a false assessment of the
dust emission. The visible absorption $A_{\rm v}$ was a free parameter of the fits. An
accurate interstellar absorption law - i.e. the wavelength
dependence of the $\frac{A_{\rm \lambda}}{A_{\rm v}}$ ratio in
the line of sight - was then needed to properly fit the SEDs. 

From 1.25~$\mu$m to 8.0~$\mu$m, we used the
analytical expression given in \citet{2005Indebetouw}. They derived it
from the measurements of the mean values of the color excess ratios
$\frac{A_{\rm \lambda}-A_{\rm J}}{A_{\rm J}-A_{\rm K}}$
from the color distributions of observed stars in the direction of the
Galactic centre. They used archival data from the GLIMPSE catalogue,
which is relevant in our case.  Beyond 8.0~$\mu$m, where absorption is dominated by the silicate
features at 9.7~$\mu$m and 18.0~$\mu$m, we used the extinction law from
\citet{1996Lutz}, which includes the interstellar silicate absorption at 9.7 $\mu$m. The $\frac{A_{\rm \lambda}}{A_{\rm v}}$ values we
used in each band are listed in Table~3.

\subsection{SEDs}

With all the observational and archival data from NIR to MIR
wavelengths, we built the 2005 and 2006 SEDs of these
sources. We fitted them ($\chi^2$ minimisation) using a
combination of two absorbed black bodies, one representing the companion
star emission and one representing a MIR excess, if there was any:
\begin{equation}
  \lambda{F(\lambda)}\,=\,\frac{2\pi{h}{c}^2}{{D_{\ast}}^2{\lambda}^4}\,10^{\textrm{\normalsize
      $-{0.4A_\lambda}$}}\left[\frac{{R_{\rm \ast}}^2}{
      e^{\textrm{\large
	  $\frac{hc}{{\lambda}k{T}_{\rm \ast}}$}}-1}+\frac{{R_{\rm D}}^2}{e^{\textrm{\large
	  $\frac{hc}{{\lambda}k{T}_{\rm D}}$}}-1}\right]\,\textrm{in W m}^{-2}
\end{equation}
We added to the data uncertainties systematic errors as following:
\begin{itemize}
\item a 2$\%$ systematic error in each IRAC band as given in the IRAC
  manual\footnote{http://ssc.spitzer.caltech.edu/documents/som/som8.0.irac.pdf},
\item comparing the variations of the flux calibration values obtained
  from the standard stars with VISIR during our observation nights, we figured
  out that systematic errors were about 5$\%$ at 10 $\mu{\textrm{m}}$ and 10$\%$ at 20 $\mu{\textrm{m}}$.
\end{itemize}
The free parameters of the fits were the absorption in the \textit{V} band $A_{\rm v}$,
the companion star black body temperature  $T_{\ast}$ and its
radius-to-distance ratio $\frac{R_{\ast}}{D_{\ast}}$, as
well as the MIR excess spherical component black body temperature
and radius $T_{\rm D}$ and $R_{\rm D}$. The best-fitting
parameters for individual sources, as well as the corresponding $\chi^2$ are
listed in Table~4, and fitted SEDs of objects A, B and C are displayed in Fig~5.

\section{Results and discussion}

\begin{figure}
  \centering
  \includegraphics[width=6.3cm,angle=270]{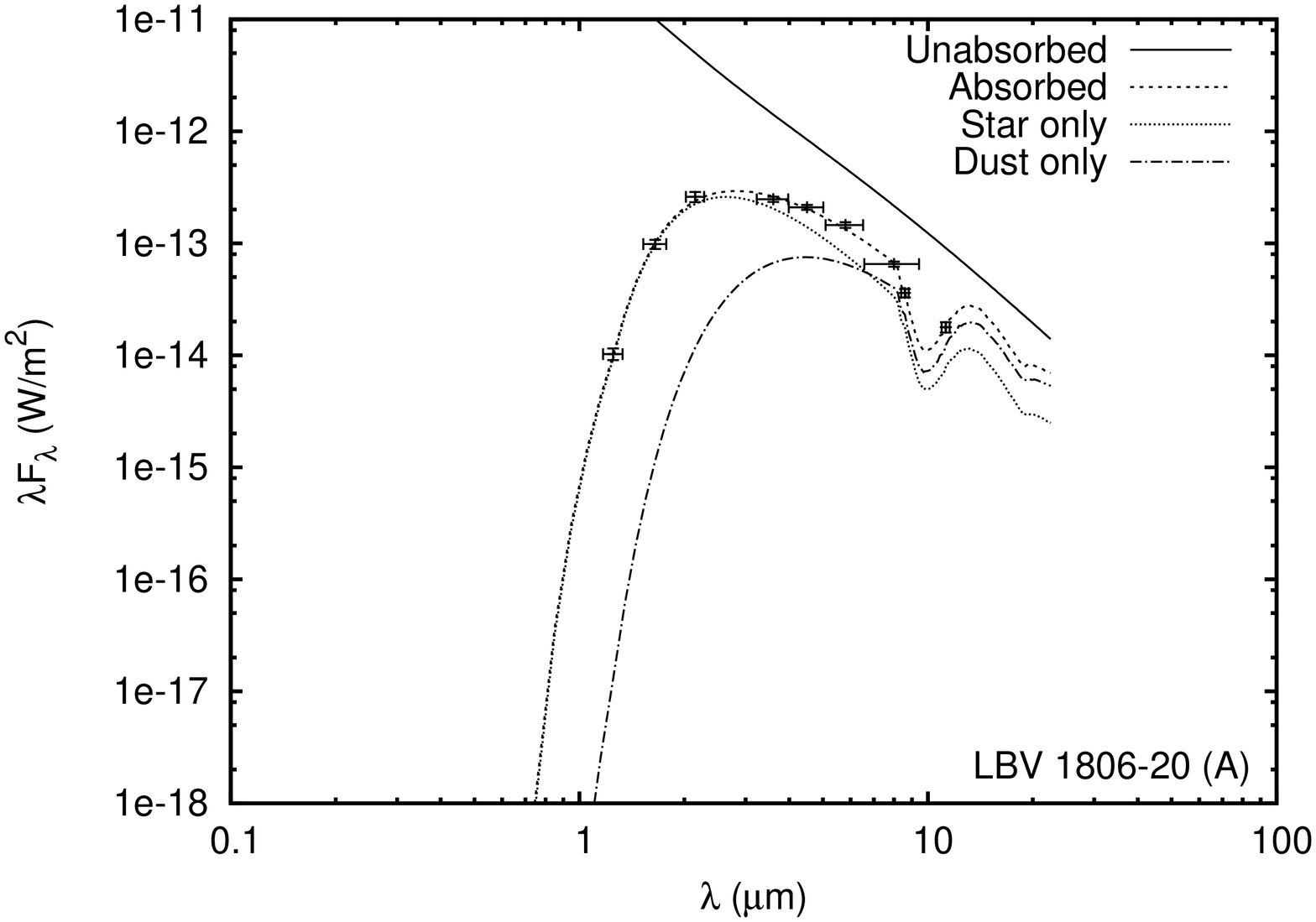}
  \includegraphics[width=6.3cm,angle=270]{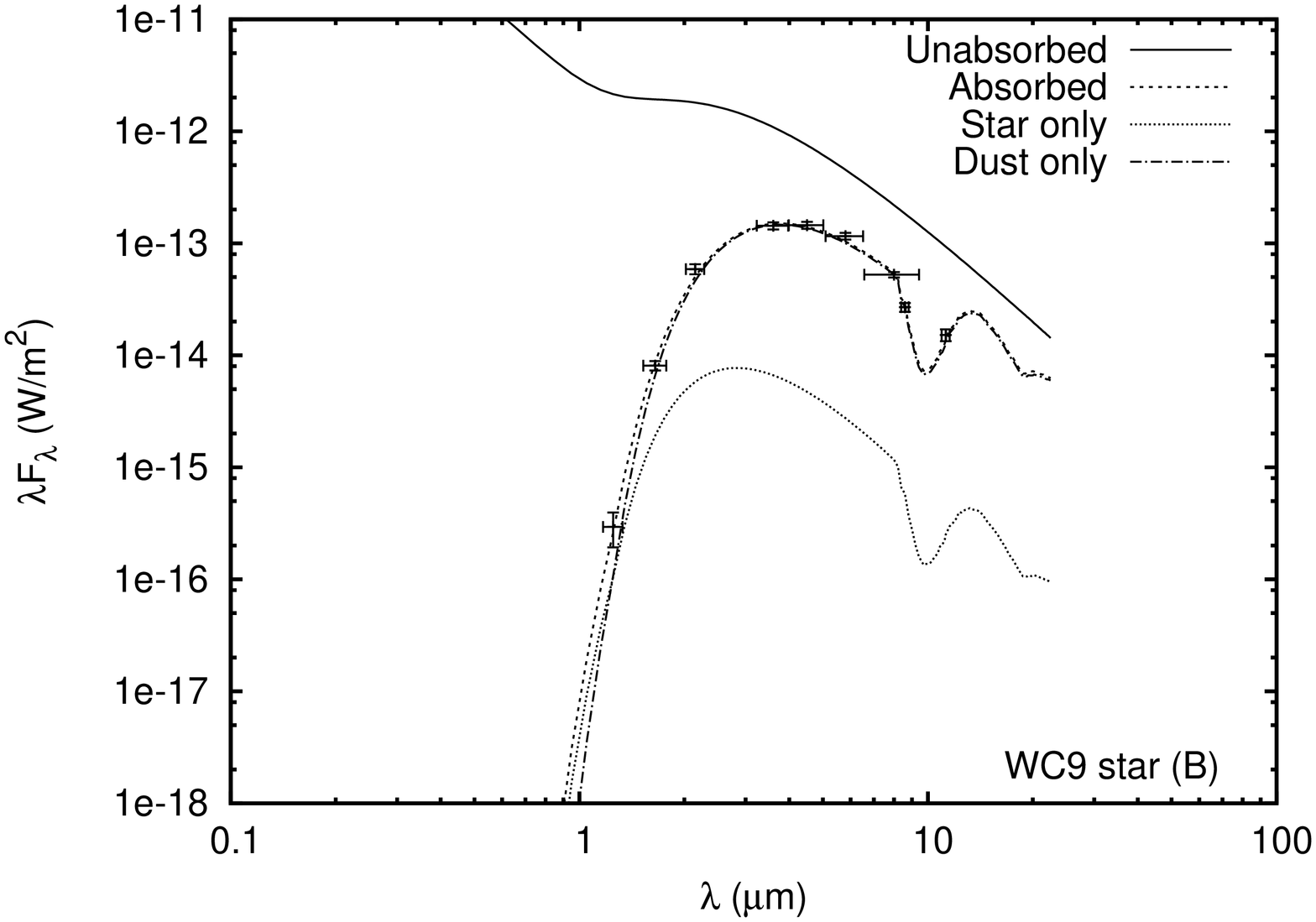}
  \includegraphics[width=6.3cm,angle=270]{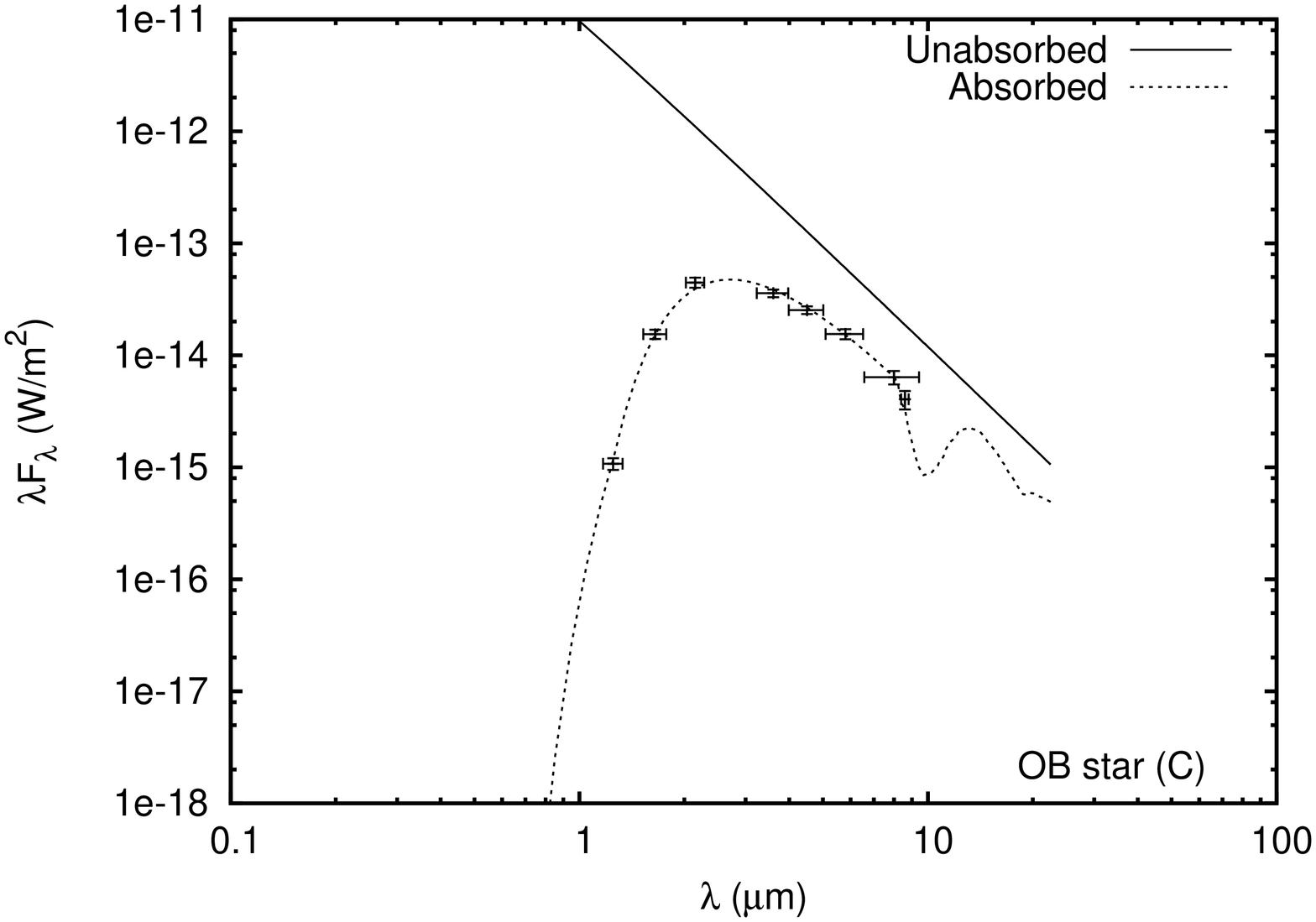}
  \caption{NIR to MIR fitted SEDs of LBV~1806-20, the WC9 star,
    and the O/B supergiant. For LBV~1806-20 and the
    WC9 star, we also add parts of their emission due to the stellar
    and the dust components. We used the MIR data from the VISIR observations we
    performed, the GLIMPSE survey, as well as the NIR data from EML04. 
    The silicate absorption feature at 9.7 $\mu$m comes from 
the use of the Lutz's interstellar extinction law, in which it is 
included.}
\end{figure}

\begin{table*}
  \caption{Adopted $\frac{A_{\rm \lambda}}{A_{\rm v}}$ values.}
  $$
  \begin{array}{c c c c c c c c c c}
    \hline
    \hline
    \textrm{Filters}&J&H&Ks&3.6\,\mu\textrm{m}&4.5\,\mu\textrm{m}&5.8\,\mu\textrm{m}&8.0\,\mu\textrm{m}
    &8.59\,\mu\textrm{m}&11.25\,\mu\textrm{m}\\
    \hline
    \frac{A_{\rm \lambda}}{A_{\rm v}}&0.289&0.174&0.115&0.064&0.054&0.047&0.044&0.060&0.061\\
    \hline
  \end{array} 
  $$
\end{table*}         
\newbox\hautbox \setbox\hautbox=\hbox{\vphantom{\rule[-0.15cm]{0cm}{0.45cm}}}
\begin{table*}
  \caption{Summary of parameters we used to fit the SEDs of the
	  sources. We give their name and then the
	  parameters themselves: the extinction in the optical
	  $A_{\rm v}$, the temperature
	  $T_{\rm \ast}$ and the
	  $\frac{R_{\rm \ast}}{D_{\rm \ast}}$ ratio of the
	  star, and the
	  temperature $T_{\rm D}$ and radius
	  $R_{\rm D}$ (in
	  $R_{\rm \ast}$ unit) of the dust component. We also add
	  the reduced $\chi^2$ we reach for each fit, as well as the
	  uncertainties on the parameters.}
  $$ 
  \begin{array}{@{\usebox{\hautbox}}ccccccc}
    \hline
    \hline
    \textrm{Sources}&A_{\rm v}&T_{\rm \ast}[\Delta{T_{\rm
          \ast}}](\textrm{K})&\frac{R_{\rm \ast}}{D_{\rm \ast}}&T_{\rm
      D} (\textrm{K})&R_{\rm D}\left(R_{\ast}\right)&\chi^2/\textrm{dof}\\
    \hline
    \textrm{LBV~1806-20 (A) (2005)}&{28.6}_{-1.5}^{+1.5}&21500\,[17900\,-\,26200]&3.08_{-0.48}^{+0.36}\times10^{-{10}}&1130_{-110}^{+130}&8.4_{-0.9}^{+1.0}&3.1/4\\
    \hline
    \textrm{LBV~1806-20 (A) (2006)}&29.0_{-1.2}^{+1.2}&17600\,[15000\,-\,21200]&3.68_{-0.36}^{+0.24}\times10^{-{10}}&1020_{-100}^{+140}&6.5_{-0.7}^{+0.7}&3.9/4\\
    \hline
    \textrm{WC9 (B)}&35.2_{-1.8}^{+1.4}&53000\,[25000\,-\,71000]&4.36_{-0.68}^{+1.36}\times10^{-{11}}&1850_{-120}^{+60}&46.0_{-14.0}^{+8.0}&6.8/4\\
    \hline
    \textrm{Supergiant (C)}&31.8_{-2.1}^{+0.8}&25700\,[11600\,-\,40500]&1.38_{-0.36}^{+0.62}\times10^{-{10}}&&&6.0/5\\
    \hline
  \end{array}
  $$ 
\end{table*}

\subsection{The variability of LBV~1806-20}

The fluxes listed in Tables~1 and 2 show that LBV~1806-20 is variable in
the MIR, in all bands from 3.6~$\mu$m to 11.25~$\mu$m. Indeed, The IRAC
fluxes substantially
increased from October 2004 to September 2005, and then decreased
from September 2005 to April 2006 to reach a similar level as in October
2004. Concerning the VISIR fluxes, we
unfortunately do not have any data taken in 2004. Nevertheless,
the same decrease in the MIR flux is observed from June 2005 to June
2006 in both PAH1 and PAH2. 

In a previous spectroscopic and photometric study of LBV~1806-20,
EML04 confirmed the likely LBV nature of this star
(van Kerkwijk et al., 1995) with spectral classes 09-B2, and showed
that there were strong
anticorrelated variations
of the equivalent widths of \ion{He}{i} 2.112~$\mu$m and
\ion{Br}{$\gamma$} lines, which led to correlated variations of the
star temperature and the number of ionising photons. Moreover,
comparison with previous photometry \citep{1995Kulkarni} showed
that the variation of temperature did not result in
the variation of the \textit{K} band magnitude of the star. As a possible
explanation, they proposed either an anticorrelated
variation of the star temperature and radius, which is typical of some
LBV stars \citep{1994Humphreys, 1996Morris}, or an
increase in the stellar wind density, resulting in more absorbed
ultraviolet continuum photons and consequently, more emission lines.

To investigate the origin of the MIR flux variation of LBV~1806-20,
and to find out whether it was related to one of the
previously described mechanisms or to the heating of the gas and dust
cloud by the high-energy emission of SGR~1806-20, we built its broadband SED using 2005 and 2006 IRAC and
VISIR data - as well as the NIR magnitudes from EML04, the
NIR flux being marginally variable - and fitted it. The best-fitting
parameters are listed in Table~4.

The presence of warm circumstellar dust around LBV stars is common \citep[see e.g.][about
  AG Car]{1988McGregor}, as they are often associated with ejected
dusty nebulae, and a component of warm dust is necessary to
explain the huge MIR excess LBV~1806-20 exhibits. 

Moreover, the fits suggest a change from a B1 ($T_{\rm \ast}=21500$~K) to a
B3 ($T_{\rm \ast}=17600$~K) spectral type, and at least a 20\% increase in the stellar radius ($R_{\rm \ast}=3.08\,\times\,
10^{-{10}}\,\times\,D_{\rm \ast}$ to $R_{\rm \ast}=3.68\,\times\,
10^{-{10}}\,\times\,D_{\rm \ast}$) from 2005 to 2006. Meanwhile, the dust temperature
decreased by about 10\% ($T_{\rm D}=1130$~K to $T_{\rm D}\,=\,1020$~K)
while the dust radius
(decrease of about 7.5\% from $R_{\rm D}=25.87\,\times\,10^{-{10}}\,\times\,D_{\rm \ast}$
to $R_{\rm D}=23.92\,\times\,10^{-{10}}\,\times\,D_{\rm \ast}$)
was roughly constant, considering the uncertainties on the star radius
and distance (about 17\%). 

Under the assumption that a dust grain is a perfect black body -
i.e. it fully absorbs the received flux - and that the star emission
is not absorbed before reaching the dust, the equation of
thermodynamical equilibrium between a dust grain and the star can be written:
\begin{equation}
  \sigma {T_{\rm \ast}^{4}} \left(\frac{R_{\rm \ast}}{R_{\rm
      D}}\right)^2\,=\,4\times \int_{0}^{\infty}{Q(\nu) B(T_{\rm D},\nu) d\nu}
\end{equation}
where $B(T_{\rm D},\nu)$ is the dust grain black body emission at the 
frequency $\nu$, and $Q(\nu)$ the dust grain emissivity, approximated
as $Q(\nu)\,=\,Q_{\rm 0}\nu^n$, where $Q_{\rm 0}$ is a constant and $1 \leq n \leq 2$ \citep{1984Draine}. 

The expected grain temperature can therefore be written:
\begin{equation}
  T_{\rm D}\,=\,\left[\left (\frac{\pi^4}{60 Q_{\rm
        0}}\right )\left(\frac{h}{k}\right)^n\frac{1}{\Gamma(4+n)\zeta(4+n)}\left(\frac{R_{\rm \ast}}{R_{\rm
    D}}\right)^2 T_{\rm \ast}^4\right]^\frac{1}{4+n}
\end{equation}
where $h$ and $k$ are the Planck and the Boltzmann constants,
respectively, $\Gamma$ the gamma function, and $\zeta$ the Riemann zeta
function.

By studying the properties of the warm dust around the LBV~HD168625,
\citet{1998Robberto} give $Q_{\rm 0}\,=\,1.52\times10^{-{8}}\times a$ -
where $a$ is the dust grain radius - and find
$a\,\sim\,1$~$\mu$m, consistent with LBVs having large dust grains ($a
\geq 0.1$~$\mu$m) in their winds, as shown in \citet{1986Mitchell} for $\eta$~Car, and
\citet{1988McGregor} and \citet{1996Shore} for AG~Car, respectively.

With $Q_{\rm 0}$ given in \citet{1998Robberto}, a
large dust grain size $a\,\sim\,0.5$~$\mu$m, and $n\,=\,1.2$ as given in \citet{1991Rouleau}, we find
$T_{\rm D}\,\sim\,1143$~K in 2005, and $T_{\rm
  D}\,\sim\,1081$~K in 2006, i.e. a decrease of about 6\%, in good
agreement with what we obtain in our best fits. This is 
therefore consistent with a MIR flux variation of LBV~1806-20 due to the anticorrelated
variations of the star temperature and radius rather than to a heating or a cooling of the gas and dust cloud due to the high-energy activity of SGR~1806-20.

\subsection{Stars B and C}

B was classified as a WC9 Wolf-Rayet (EML04), and the fluxes listed in
Table~1 and 2 show that its MIR emission was constant in all bands
from 3.6~$\mu$m to 11.25~$\mu$m. 
We then fitted its SED using the MIR fluxes of 2006 and the NIR magnitudes found in EML04, and the best-fitting parameters are listed in Table~4. Our results suggest that B is extremely
reddened and that it exhibits a very large MIR excess. The presence of such
excess is typical of WC9 stars, whose C rich stellar winds allow the creation of
circumstellar dust. Moreover, Fig~5 shows that beyond 1.6~$\mu$m (\textit{H}
band), the star barely contributes to the emission,
which is also common for WC9 stars \citep{1996Hucht, 2006Crowther2}. All the fits with a more balanced contribution of the stellar and
the dust components (reduced $\chi^2\,\ge\,3$), or with a 
stellar component only (reduced $\chi^2\,\ge\,100$) failed to reproduce the
SED. Moreover, we point out that although $T_{\rm D}=1850\,\textrm{K}$ is high, it is still below the sublimation temperature of amorphous carbon found
around WC9 stars \citep[$\sim$2000~K, see for instance][]{1993Laor,
  1993Preibisch}.
\newline

C was found to be either an O/B supergiant or an hypergiant through
spectroscopy (EML04), then \citet{2005Figer} performed higher resolution
spectroscopy and derived an O/BI spectral class, with narrower lines
than previously found. Their result is confirmed by \citet{2008Bibby} who derive 
a B1$-$B3I spectral type through high-resolution NIR spectroscopy. 
We unfortunately did not detect C during our first VISIR run in June
2005 as our exposure time was not sufficient, preventing us to
perform a comparison with
the flux of June 2006. Nevertheless, the IRAC fluxes listed in Table~2
point out towards a constant MIR flux from 2004 to 2006 in all
bands. Using the NIR magnitudes given in EML04, as well as the IRAC
and the VISIR fluxes of 2006, we fitted its SED and the best-fitting parameters are
listed in Table~4.

The best-fitting temperature is typical of an early$-$B supergiant
star. Moreover, if we assume that the
star C is associated with the cluster, it is then possible to get its radius by
multiplying the best-fitting radius-to-distance ratio derived from our
fit by the distance of the cluster. If we use a distance of $15.1_{-1.3}^{+1.8}$~kpc given in \citet{2004Corbel}, 
we derive a radius $R_{\rm \ast}\,=\,92.4_{-32.1}^{+53.2}\,\textrm{R}_\odot$, while we find 
$R_{\rm \ast}\,=\,53.2_{-18.5}^{+30.6}\,\textrm{R}_\odot$ for a distance of $8.7_{-1.5}^{+1.8}$~kpc \citep{2008Bibby}. The latter 
is more consistent with the expected radius of a normal
B1$-$B3 supergiant star, therefore favouring the distance given 
in \citet{2008Bibby}. 

Finally, we would like to point out that the three-color image
displayed in Fig~2 interestingly suggests that C is associated with
the northern part of the gas and dust cloud, since the cloud emission reappears at the supergiant position. If this were confirmed, it would mean that the cloud is heated by C.

\subsection{SGR~1806-20}

High-energy emission of magnetars is believed to be powered by the
magnetic energy of their very strong magnetic field. Moreover, five AXPs and SGR~1806-20
have been detected in the optical bands and/or the NIR, and all of
them were found to exhibit a variable NIR emission correlated to the
X-ray emission. This optical/NIR emission is
explained either as the non-thermal radiation by particles in the
magnetosphere, or as the irradiation by the X-ray emission
of a fossile passive dust disk around the neutron star.

The second explanation recently received some credit as the AXP
4U~0142+61 has been detected in the MIR using IRAC at 4.5~$\mu$m and 8
$\mu$m \citep{2006Wang}. It is shown that its IR
emission can be understood by the presence of an X-ray heated
disk. This disk would have formed from the fallback material of a supernova, and
would emit mostly in the IR consequently to the irradiation by the
high-energy emission of the neutron star. Moreover, the authors derive an
unabsorbed MIR to X-ray flux ratio
$\nu_{\rm 4.5\mu{m}}F_{\rm 4.5\mu{m}}/F_{\rm X}\,\sim\,3.4\times10^{-{4}}$, where
F$_{\rm X}$ is the unabsorbed 2-10~keV X-ray flux.
\newline

SGR~1806-20 was never detected in the MIR domain, neither with IRAC nor 
with VISIR, as shown in the images displayed in this paper. Indeed, considering
that the MIR emission of SGR~1806-20 would be due to an irradiated fallback
disk, and that the unabsorbed MIR to X-ray flux ratio derived
in \citet{2006Wang} is typical of dust disks around young neutron
stars, we can expect the SGR~1806-20 absorbed flux at 4.5~$\mu$m to be
below 10~$\mu$Jy, one order of magnitude below the sensitivity of
about 200~$\mu$Jy at 5$\sigma$ of the GLIMPSE survey.

\subsection{Absorption}

All the intrinsic absorptions derived from our fits 
show that stars B and C are likely more absorbed than LBV~1806-20,
as they are embedded in a hotter and denser part of the associated cloud of dust and gas visible on our VISIR images in PAH2 and Q2 (see Fig~2). Indeed, Fig~4 displays a MIPS image of the
environment of SGR~1806-20 at 24~$\mu$m with contour plots. The \textit{Spitzer}
sensitivity being far better, a larger part of the cloud is
visible. We see that LBV~1806-20 (A) is located in a colder and less
dense zone than the WC9 star (B) and the B supergiant star (C), which
both are closer to the core of the cloud. Note that the image is
saturated, and the saturated pixels correspond to the zone of the
cloud detected with VISIR (taking into account the difference
of plate scale per pixel in both images, 0\farcs075 for VISIR and 2\farcs5
for MIPS). B is then likely very close to the hottest and densest
zone of the dust cloud, which is why it exhibits the highest
intrinsic absorption. We nevertheless point out that a limitation to
this conclusion is that we only see the cloud integrated along the line of
sight.  

\section{Conclusions}

We reported the mid-infrared photometry of three stars and the gas
and dust cloud associated with the
same massive star cluster as SGR~1806-20, using ESO/VISIR and \textit{Spitzer}/IRAC-MIPS data
obtained at different epochs between 2004 and 2006. 

We show that LBV~1806-20 is the only object to exhibit a likely MIR
variability, and that this flux variation is probably the consequence
of its LBV nature rather than a heating of its circumstellar
dust by the giant flare exhibited by SGR~1806-20. We also show that
the stars in the central zone of the massive star cluster appear more
absorbed, as they are closer to the hottest and densest part of the gas
and dust cloud in which all of them are embedded. 

Finally, we recommend further high-sensitivity and long
exposure MIR observations of SGR~1806-20 in order to try to detect it, 
and perhaps constrain the origin of the AXPs and SGRs optical/IR emission. 

%______________________________________________________________

\begin{acknowledgements}

  We are pleased to thank the anonymous referee for useful comments, 
  as well as E. Pantin for his IDL code - part of the
  ESO/VISIR reduction pipeline - we used to
  reduce our VISIR data, and J.L. Starck for his wavelets
  package we used to clean our images. FR acknowledges the CNRS/INSU
  for the funding of the third year of his ESO/CEA studentship. This
  research has made
  use of NASA's Astrophysics
  Data System, of the SIMBAD and VizieR databases operated at
  CDS, Strasbourg, France, of products from, as well as products from
  the Galactic Legacy Infrared Mid-Plane Survey
  Extraordinaire, which is a \textit{Spitzer Space Telescope}
  Legacy Science Program.

\end{acknowledgements}
\bibliographystyle{aa}
\bibliography{./mybib}{}

\end{document}